\documentclass[letterpaper, 10 pt, conference]{ieeeconf}
\IEEEoverridecommandlockouts
\usepackage{cite}
\usepackage{amsmath,amssymb,amsfonts}
\usepackage{graphicx}
\usepackage{textcomp}
\usepackage{xcolor}

\newcommand{\algrule}[1][.2pt]{\par\vskip.5\baselineskip\hrule height #1\par\vskip.5\baselineskip}

\usepackage{xspace}
\usepackage{soul}
\usepackage{mathtools}
\usepackage{makecell}
\usepackage{cite}
\usepackage{nicefrac}
\usepackage{float}
\usepackage{wrapfig}
\usepackage{amssymb}
\usepackage{amsmath}
\usepackage{xcolor}
\usepackage{graphicx}

\usepackage{subcaption}
\usepackage[]{algorithm2e}
\usepackage{tikz}
\usepackage{pgfplots}
\pgfplotsset{compat = 1.3}

\newcommand{\ra}{\rightarrow}

\newtheorem{asm}{Assumption}
\newtheorem{prp}{Proposition}
\newtheorem{thm}{Theorem}

\newtheorem{rem}{Remark}

\newcommand{\ie}{\unskip, i.\,e.,\xspace}
\newcommand{\eg}{\unskip, e.\,g.,\xspace}

\newcommand{\sut}{\text{s.\,t.\,}}
\newcommand{\wrt}{w.\,r.\,t.\xspace}

\newcommand{\N}{\ensuremath{\mathbb{N}}}

\newcommand{\R}{\ensuremath{\mathbb{R}}}

\newcommand{\V}{\ensuremath{\mathcal{V}}}

\newcommand{\X}{\ensuremath{\mathbb{X}}}

\newcommand{\U}{\ensuremath{\mathbb{U}}}

\newcommand{\eps}{\ensuremath{\varepsilon}}

\newcommand{\ball}{\ensuremath{\mathcal B}}
\newcommand{\K}{\ensuremath{\mathcal{K}}\xspace}		

\DeclareMathOperator*{\arginf}{arg\,inf}

\makeatletter
\newcommand{\subalign}[1]{%
	\vcenter{%
		\Let@ \restore@math@cr \default@tag
		\baselineskip\fontdimen10 \scriptfont\tw@
		\advance\baselineskip\fontdimen12 \scriptfont\tw@
		\lineskip\thr@@\fontdimen8 \scriptfont\thr@@
		\lineskiplimit\lineskip
		\ialign{\hfil$\m@th\scriptstyle##$&$\m@th\scriptstyle{}##$\crcr
			#1\crcr
		}%
	}
}

\usepackage{ marvosym }

\title{\LARGE \bf
	Approximate infinite-horizon predictive control}

\author{Lukas Beckenbach
	and~Stefan~Streif
	\thanks{Lukas Beckenbach
		and~Stefan~Streif are with the Laboratory for Automatic Control and System Dynamics, Technische Universität Chemnitz, 09107 Chemnitz, Germany. {\tt\footnotesize \{lukas.beckenbach,stefan.streif\}@etit.tu-chemnitz.de}}
}

\begin{document}
	
	\maketitle
	\thispagestyle{empty}
	\pagestyle{empty}

%

\maketitle

\begin{abstract}
Predictive control is frequently used for control problems involving constraints. 
Being an optimization based technique utilizing a user specified so-called stage cost, performance properties \ie bounds on the infinite horizon accumulated stage cost, aside closed-loop stability are of interest. 
To achieve good performance and to influence the region of attraction associated with the prediction horizon, the terminal cost of the predictive controller's optimization objective is a key design factor. 
Approximate dynamic programming refers to one particular approximation paradigm that pursues iterative cost adaptation over a state domain. 
Troubled by approximation errors, the associated approximate optimal controller is, in general, not necessarily stabilizing nor is its performance quantifiable on the entire approximation domain. 
Using a parametric terminal cost trained via approximate dynamic programming, a stabilizing predictive controller is proposed whose performance can directly be related to cost approximation errors. 
The controller further ensures closed-loop asymptotic stability beyond the training domain of the approximate optimal controller associated to the terminal cost. 
\end{abstract}


\section{Introduction}
\label{sec:introduction}

Predictive control is a particular control design involving repetitive solving an optimal control problem \cite{Gruene2017}. 
It requires a system model for predicting the system's state, which then is optimized by minimizing an objective function subject to constraints and finding the minimizing input sequence. 
Inclusion of constraints constitute a major benefit of the approach. 
The objective function is commonly constructed using a finite accumulation of a user-defined stage cost and potentially includes a terminal cost penalizing the terminal predicted state. 
The former can further be used to define an infinite-horizon optimal control problem, in which an infinite accumulation of stage cost should be minimized by an input sequence. 
Due to computational difficulties arising from the so-called curse of dimensionality, only suboptimal controllers with respect to the infinite-horizon cost are available in general \cite{Lincoln2006}. 
Such suboptimal control strategies include, for instance, predictive control in which the suboptimality is induced by finite truncation of the infinite-horizon problem. 
To reduce the suboptimality, terminal costs can be used to approximate the optimal infinite-horizon tail \ie the optimal cost from the terminal predicted state onward. 
This terminal cost can be computed offline by various strategies which inherit different approximation qualities and thus optimality properties. 

Quasi-infinite-horizon predictive control \cite{Chen1998} refers to such ideology, finding in particular local approximations to the optimal cost in a vicinity of the origin, the terminal set. 
For constructing approximate optimal infinite-horizon tails, common approaches include \eg local linear quadratic regulators \cite{Chen1998,Lazar2018}, Al'Brehkts method \cite{Krener2021a,Lucia2015a} or finite-horizon tail \cite{Koehler2021,Magni2001} extension under known local control laws. 
Asymptotic stabilization is then guaranteed in dependence of the approximation properties further assuming either terminal constraints, leading the terminal predicted state to the terminal set, or by a sufficiently long prediction horizon. 

Approximate dynamic programming \cite{Lewis2009,Zhang2013} refers to an iterative adaptation scheme capable of approximating the optimal infinite-horizon cost and the associated controller over a state domain. 
This is commonly done using parametric architectures like \eg neural nets, and performing an approximation iteration over state samples prior to the actual control process which may be performed on the trained controller. 
Among several manifestations of this scheme, some specifically consider parametric approximation structures and allow analyzing the effect of approximation errors on convergence of the approximate cost and stability of the resulting closed-loop under the iterated controller, respectively \cite{Heydari2018a,Liu2013}. 
Contrary to other cost approximation techniques, they are designed for data-based approximation without requiring system assumptions such as, for instance, smoothness or stabilizability of the linearized dynamics as some of the others.  

In the view of increasing interest in learning based methods in predictive control, a predictive controller terminal cost construction under an approximate dynamic programming scheme stands to reason. 
Although specifically aiming at the problem of infinite-horizon optimality, the use of iteratively adapted approximate cost of ADP as optimal infinite-horizon tail approximation has not gained much attention. 
However, the approximation can be computed prior to the control process and stability and performance of the predictive controller can be deduced by the choice of the prediction horizon and the used parametric architecture.  

This work proposes an approximate infinite-horizon predictive controller by approximating the optimal infinite-horizon tail locally around the origin using a particular approximate dynamic programming scheme, introduced in Section~\ref{sec:problem-setup-preliminaries}. 
Using specific properties of this scheme, a prediction horizon is deduced in Section~\ref{sec:analysis-main} that renders the closed-loop asymptotically stable and permits a performance statement on the predictive controller. 
Explicit dependence of the performance with respect to errors of the cost approximation is further shown. 
A case study in Section \ref{sec:case-study} depicts the influence of the approximation errors on the performance of the predictive controller.


\section{Problem Setup and Preliminaries}
\label{sec:problem-setup-preliminaries}

This work considers discrete-time systems of the form
\begin{align}
	\label{eq:system}
	x(k+1) = f(x(k),u(k)) \coloneqq f_a(x(k)) + g_a(x(k))u(k)
\end{align}
for continuous $f_a:\X \ra \R^n$, $g_a:\X \ra \R^{n \times m}$, satisfying $f_a(0)= g_a(0) =0$, and initial conditions $x(0) \in \X$, in which $\X \subset \R^n$ and $\U \subset \R^m$ denote compact state $x \in \X$ and input $u \in \U$ constraints, respectively. 
Let $\{0\}$ be in the interior of $\X$ and $\U$ and denote $\U^N$ a length $N \in \N$ sequence of control inputs $u(k)$ such that the associated state satisfies $x(k) \in \X$ along \eqref{eq:system} for all $k \in \{1,\dotsc,N\}$ staring at $x(0) \in \X$. 
Further, it assumes a continuous used-defined positive definite stage cost $l: \X \times \U \ra \R_{\geq 0} $ \ie $l(0,0) = 0$ and $l(x,u)>0$ else, of the form $l(x,u) = Q(x) + u^\top R u$, $Q:\X \ra \R_{\geq 0}$ positive definite and $R = R^\top \succ 0$, which is used to construct the infinite-horizon cost
\begin{align}
	J(x(0),u(\cdot)) \coloneqq \sum_{k=0}^{\infty} l(x(k),u(k)),
\end{align}
whose minimum over an infinite-horizon sequence $u(\cdot) \in \U^{\infty}$ is denoted by $V$ \ie for any $x(0) \in \X$,
\begin{align}
	V(x(0)) \coloneqq \min_{u(\cdot) \in \U^{\infty}}  \; J(x(0),u(\cdot)).
\end{align}
For convenience, denote $l^\ast(x) \coloneqq \min_{u\in \U} l(x,u) = Q(x)$ on which the following is implied (refer to \eg \cite{Boccia2014}). 
\begin{asm} \label{asm:stage-cost-lower-bound}
	There exist $\mathcal{K}$-functions $\underline{\alpha}_l,\overline{\alpha}_l$ \sut $\underline{\alpha}_l(\|x\|) \leq l^\ast(x) \leq \overline{\alpha}_l(\|x\|)$ for all $x \in \X$. 
\end{asm}
For a definition of $\mathcal{K}$-functions, the reader is referred to \cite{Khalil2002}. 
Under a suitable assumption, for instance the controllability Assumption \ref{asm:loc-exp-controllability-iter-ctrl} introduced later, it can be shown that $V$ is finite on subsets of $\X$ (see \eg \cite{Gruene2017}). 

Consider a bounded $\Omega \subset \X$ with the origin in its interior and a continuous, positive-definite $V_f:\Omega \ra \R_{\geq 0}$. 
For a fixed horizon length $N \in \N$, the herein investigated predictive controller's optimization problem is given by
\begin{align} \label{eq:MPC-OCP}
	\begin{split}
		\V_N(x) \coloneqq \hspace{-0.5em}\min_{u(\cdot) \in \U^N} &\sum_{k=0}^{N-1} l(x(k),u(k)) + V_f(x(N)) \\
		\text{\sut} \; \;  &x(k+1) = f(x(k),u(k)), \; x(0) = x, \\
		& x(k) \in \X, \; k= 0,\dotsc,N,\\
		&u(k) \in \U, \; k= 0,\dotsc,N-1,
	\end{split}
\end{align}
to which $\kappa_{N}(x)$ is the first element of the minimizing sequence and which is solved at each state $x = x(k;\kappa_{N})$ along the resulting closed-loop $x(k+1;\kappa_{N}) = f(x(k;\kappa_{N}),\kappa_{N}(x(k;\kappa_{N})))$ with $ x(0;\kappa_{N})= x(0)$. 
Considering the domain of $V_f$, it needs to be guaranteed that $x(N) \in \Omega$.  
\begin{rem}
	The latter is commonly achieved under a sufficiently long prediction horizon in dependence of properties of the terminal cost. 
	For instance, using $V_f$ satisfying \eg $\min_{u \in \U} \{l(x,u) + V_f(f(x,u)) - V_f(x) \}\leq 0$, for all $x \in \Omega$, allows exploiting monotonicity properties of the optimal cost and thereby deduce a minimal horizon length \cite{Limon2006}. 
	Contrary, so-called relaxed local Lyapunov terminal costs as in \eg \cite{Grimm2005,Tuna2006} requires a more elaborate horizon derivation. 
\end{rem}

The prediction horizon is selected such that the relaxed dynamic programming inequality holds (see \eg \cite{Lincoln2006}) \ie 
\begin{align} \label{eq:RDP}
	\V_N(f(x,\kappa_{N}(x))) + \alpha l(x,\kappa_{N}(x)) \leq \V_N(x)
\end{align}
for a $\alpha \in (0,1]$ and all $x \in \X$, from which asymptotic stability is concluded. 
Aside relating $\V_N$ to $V$, this work characterizes $\alpha$ in dependence of $N$ as well as the properties of the terminal cost, adapted by approximate dynamic programming and introduced next. 

\subsection{Approximate Dynamic Programming}
\label{subsec:ADP}

Approximate dynamic programming has been regarded as a subfield of reinforcement learning and can be used to iteratively adapt an approximant $V_{i}(x)$ to the optimal cost $V(x)$ on some $\Omega \subset \X$. 
Approximation on $\Omega$ is done by employing parameteric structures, such as single layer neural networks, for $V^{i}$. 
Existing iteration schemes can roughly be classified as those adapting to $V$ from below (refer to \eg \cite{Al-Tamimi2008-VI-ADP} and the notion of value iteration), starting at zero initial guess $V_{0}(x) \equiv 0 $, or those approximating the optimal cost from above associating $V_{0}$ with an asymptotically stabilizing control law (see \eg policy iteration \cite{Liu2014}). 
\emph{Stabilizing} value iteration also corresponds to the latter case and uses a known control law $\mu_{-1}:\Omega \ra \R^m$ upon which iterative controllers and cost functions are derived performing
\begin{enumerate}
	\item[i)] an initial cost approximation by solving
	\begin{align} \label{eq:ADP-init-cost}
		\begin{split}
		&\mathcal{L}_0(V_{0}) \coloneqq \\
		&V_{0}(x) - l(x,\mu_{-1}(x)) - V_{0}(f(x,\mu_{-1}(x))) =0,
		\end{split}
	\end{align}

	\item[ii)] and for each $i = 0,1,\dotsc$, obtaining an iterative control law by letting
	\begin{align} \label{eq:ADP-iter-ctrl}
		\mu_{i}(x) = \arginf_{u \in \R^m} \, \{ \, l(x,u) + V_{i}(f(x,u)) \, \}, 
	\end{align}
	
	\item[ii)] as well as updating the approximate cost by
	\begin{align} \label{eq:ADP-iter-cost}
		\begin{split}
			&\mathcal{L}_i(V_{i+1},V_{i}) \coloneqq \\
			&V_{i+1}(x) - l(x,\mu_{i}(x)) - V_{i}(f(x,\mu_{i}(x))) =0,
		\end{split}
	\end{align}
\end{enumerate}
for all $x \in \Omega$. 
It has been shown in \eg \cite{Heydari2016} that $V_{i}(x) \ra V(x)$ for $i\ra \infty$ monotonically from above. 
\begin{rem}
	It may be noteworthy that although $\mu_{i}$ can be expressed analytically by the first-order condition, 
	certain discrepancy between $\lim_{i \ra \infty}V_{i}(x)$ and $V(x)$ may arise, in general, due to negligence of constraints. 
	It may be argued however that under conditions such as \eg the small control property, control constraints are inactive in a vicinity of the origin.
\end{rem}
\begin{asm} \label{asm:ADP-init-cost-bound}
	There exist $\gamma_0 >0$ \sut $V_{0}(x) \leq \gamma_0 l^\ast(x)$ for all $x \in \Omega$. 
\end{asm}
Unfortunately the above equations are rarely tractable \ie solvable for all $x \in \Omega$, and once parametric structures such as 
\begin{align}
	\hat{V}_{i}(x) \coloneqq w^\top(i) \Phi(x),
\end{align}
for some continuous basis function vector $\Phi:\Omega \ra \R^l$ and weights $w(i) \in \R^l$, for all $i \in \N_0$, are used to embody the auxiliary approximant $V_{i}$, only relaxed solutions can be obtained in the form of 
\begin{subequations} \label{eq:ADP-AVI-errors}
	\begin{align}
		\mathcal{L}_0(\hat{V}_{0}) &= \eps_{-1}(x)  \label{eq:ADP-AVI-errors-1}\\
		\mathcal{L}_i(\hat{V}_{i+1},\hat{V}_{i}) &= \eps_{i}(x), \label{eq:ADP-AVI-errors-2}
	\end{align}
\end{subequations}
for some $\eps_{i}:\Omega \ra \R$, for all $i \in \N_0 \cup \{-1\}$ and all $x \in \Omega$. 
\begin{rem}
	Errors in the equations may also arise due to the fact that instead of the entire $\Omega$, a set of sample points $\hat{\Omega} \subset \Omega$ is used by learning based approaches to train $\hat{V}_i$. 
	Hence, though the equations may be fulfilled at the samples of $\hat{\Omega}$, offsets may be expected at inter sample states which are covered by the error $\eps_i$. 
	Subsequently, the error margin for another set of testing samples inside $\Omega$ must be inspected.  
\end{rem}
The resulting scheme is referred to as stabilizing value iteration with errors (AVI). 
A bound on the error $\eps_{i}$ used in \cite{Heydari2016b} reads as follows.
\begin{asm} \label{asm:ADP-error-bound}
	There exists $c \in [0,1)$ \sut $|\eps_{i}(x)| \leq c \, l^\ast(x)$ for all $x \in \Omega$ and $i \in \N_0 \cup \{-1\}$. 
\end{asm}
Sufficiently small errors may be accomplished \eg by increasing, loosely speaking, the richness of basis functions, that is, increasing $l$ with suitable selection of $\Phi$ (refer to \eg \cite{Abu-Khalaf2005}). 
An error bound can be computed at each iteration $i$ by testing \eqref{eq:ADP-AVI-errors} for representative \ie sufficiently many test states in $\Omega$ (beyond the samples $\hat{\Omega}$).

\begin{thm}[from \cite{Heydari2018a}] \label{thm:ADP-AVI-Heydari}
	Let $\hat{V}_{i}$ be obtained from \eqref{eq:ADP-AVI-errors} and $\mu_{i}$ from \eqref{eq:ADP-iter-ctrl} using $\hat{V}_{i}$. 
	Let Assumptions \ref{asm:stage-cost-lower-bound}--\ref{asm:ADP-error-bound} hold. 
	Then the approximate cost satisfies
	\begin{enumerate}
		\item[i)] the lower bound $(1-c) l^\ast (x) \leq \hat{V}_i(x)$,
		\item[ii)] the difference inequality
		\begin{align} \label{eq:Viter-difference-Heydari}
			\begin{split}
				&\hat{V}_i(f(x,\mu_{i}(x))) - \hat{V}_i(x) \\
				\leq\;  &- l (x,\mu_{i}(x)) - \eps_i(x) + \frac{4c}{1-c}V_0(x),
			\end{split}
		\end{align}
	\item[iii)] and the upper bound $\hat{V}_i(x) \leq 2 \gamma_0 l^\ast(x)$,
	\end{enumerate}
	for all $x \in \Omega$ and all $i \in \N_0$. 
	Suppose that further
	\begin{align} \label{eq:AVI-error-bound-stability}
		0 \leq c < 1 +2 \gamma_0 - \sqrt{4 \gamma_0^2 + 4 \gamma_0},
	\end{align}
	then $\mu_{i}(\cdot)$ locally asymptotically stabilizes the system and $\hat{\ball}_r^i \coloneqq \{x \in \X: \, \hat{V}_{i}(x) \leq r\}$ with $r>0 $ \sut $\hat{\ball}_r^i \subset \Omega$ is a region of attraction. 
\end{thm}

Though if \eqref{eq:AVI-error-bound-stability} were satisfied, an approximate optimal controller on $\hat{\ball}_r^i$ could be established, constraint satisfaction as well as the region of attraction may be unsatisfactory (and the controller's performance $J(x,\mu_{i}(\cdot))$ is questionable since $\hat{V}_{i+1}(x)$ approximates the optimal infinite-horizon cost).  

Next, a discussion on $\Omega$ and its impact in the predictive controller \eqref{eq:MPC-OCP} is made when $V_f$ in \eqref{eq:MPC-OCP} is replaced by an offline trained approximant $\hat{V}_i$ under AVI. 
Suppose the above iteration is performed until some fixed $i = \mathcal{I} \in \N_0$. 
Denote $\hat{\mu}(x) \coloneqq \hat{\mu}_{\mathcal{I}}(x)$ and $\hat{V}(x)\coloneqq \hat{V}_{\mathcal{I} + 1}$ for brevity. 
The subsequent analysis assumes the terminal cost in \eqref{eq:MPC-OCP} is substituted by the trained approximate optimal cost \ie $V_f(x) \equiv \hat{V}(x)$. 

\subsection{Preliminary Local Analysis}

Suppose the local controller $\mu_{-1}$ does satisfy the following controllability condition. 
\begin{asm} \label{asm:loc-exp-controllability-iter-ctrl}
	For any $M \in \N_0$, there exists $C \geq 1$ and $\sigma \in (0,1)$ \sut 
	\begin{align}
		l(x(k),\mu_{-1}(x(k))) \leq C \sigma^k l^\ast(x(0))
	\end{align}
	along $x(k+1) = f(x(k),\mu_{-1}(x(k)))$, for any $x(0) \in \Omega$ and $k = 0,\dotsc,M$. 
	Furthermore, $x(k) \in \X$ and $\mu_{-1}(x(k)) \in \U$. 
\end{asm}
For feasibility purposes, suppose further that the iterated controller does satisfy input constraints for states in $\Omega$. 
\begin{asm} \label{asm:ctrl-iter-input-constraint-satis}
	For all $x \in \Omega$, $\hat{\mu}(x) \in \U$. 
\end{asm}
A sufficiently small $\Omega$ may be selected to satisfy this requirement assuming the systems entails \eg a small control property as mentioned previously.

It is tempting to consider $\hat{\ball}_r^{\mathcal{I}}$ for the forthcoming local analysis. 
Choosing 
\begin{align} \label{eq:term-set}
	\X_f \coloneqq \{x \in \X: \, l^\ast(x) \leq \frac{d}{2 \gamma_0 C}\}
\end{align}
with $d>0$ such that $\{x \in \X: \, l^\ast(x) \leq d/(2 \gamma_0)\} \subset \Omega$ however has certain benefits as follows. 
\begin{prp} \label{prp:local-cost-bound}
	Let Assumptions \ref{asm:stage-cost-lower-bound}--\ref{asm:loc-exp-controllability-iter-ctrl} hold. 
	Then for all $x \in \X_f$, 
	\begin{align}
		\V_N(x) \leq \alpha_{\V}(\|x\|)
	\end{align}
	for some $\alpha_{\V} \in \K$, for all $N \in \N$.  
\end{prp}
\begin{proof}
	Given $\X_f$ as in \eqref{eq:term-set}, it holds that 
	\begin{align*}
		\frac{d}{2\gamma_0} \geq\; &C l^\ast(x(0))  \geq C \sigma^k l^\ast(x(0))\geq l(x(k),\mu_{-1}(x(k))) \\
		\geq \; &l^\ast(x(k))
	\end{align*}
	for any $x(0) \in \X_f$, any $k \in \{0,\dotsc,M\}$ with $M $ of Assumption \ref{asm:loc-exp-controllability-iter-ctrl}, along $x(k+1) = f(x(k),\mu_{-1}(x(k)))$. 
	Subsequently, $x(k) \in \Omega$. 
	By optimality, 
	\begin{align*}
		\V_N(x(0)) \leq \sum_{k=0}^{N-1} l(x(k),\mu_{-1}(x(k))) + V_f(x(N))
	\end{align*}
	since $(x(k),\mu_{-1}(x(k))) \in \X \times \U$ constitute feasible pairs by Assumption \ref{asm:loc-exp-controllability-iter-ctrl}. 
	Because $x(N) \in \Omega$ also, $V_f(x(N)) = \hat{V}(x(N)) \leq 2 \gamma_0 l^\ast(x(N)) \leq 2 \gamma_0 l(x(N),\mu_{-1}(x(N)))$, and thus it further holds that
	\begin{align} \label{eq:MPC-cost-local-bound}
		\begin{split}
		\V_N(x(0))\hspace*{-2pt} \leq\hspace*{-4pt} &\sum_{k=0}^{N-1}\hspace*{-3pt} l(x(k),\hspace*{-1pt}\mu_{-1}(x(k))) \hspace*{-2pt}+ \hspace*{-2pt}2 \gamma_0 l(x(N),\hspace*{-1pt}\mu_{-1}(x(N))) \\
		\leq &C \frac{1-\sigma^N}{1- \sigma} l^\ast(x(0)) + 2 \gamma_0 C \sigma^N l^\ast(x(0)) \\
		\leq & \underbrace{C \left( \frac{1}{1- \sigma} + 2 \gamma_0 \right)}_{\eqqcolon \gamma_{\V}} \; \underbrace{l^\ast(x(0))}_{ \leq \overline{\alpha}_l(\|x(0)\|)},
\end{split}	
\end{align}
	 for all $x(0) \in \X_f$.
\end{proof}

By considering $\X_f$ of \eqref{eq:term-set} as opposed to $\hat{\ball}_r^{\mathcal{I}}$ (as defined in Thm.~\ref{thm:ADP-AVI-Heydari}), set membership $x(k) \in \Omega$ -- as required for the local bound and the use of properties of AVI -- can be provided straightforward. 
It further allows to use known analyses, specifically that of \cite{Koehler2021}, for deriving desirable prediction horizons. 

\begin{rem}
	A bound as per Assumption~\ref{asm:ADP-init-cost-bound} reflects a controllability type property which can be found frequently in the predictive control literature (see \eg \cite{Gruene2008}). 
	The bound \eqref{eq:MPC-cost-local-bound} is a conservative estimate considering the fact that $\hat{V}_i$ is an estimate of $V_{\infty}$ which, under the preceding Assumption~\ref{asm:loc-exp-controllability-iter-ctrl}, can be upper bounded as per $V_{\infty}(x) \leq \gamma l^\ast(x)$, $\gamma = C/(1-\sigma)$ -- tighter than $\gamma_{\V}$.  
	Since under (nominal) stabilizing value iteration as per \eqref{eq:ADP-init-cost}--\eqref{eq:ADP-iter-cost} it also follows that $V_0(x) \geq V_i(x) \geq V_{\infty}(x)$ (see \cite{Heydari2016}), a potentially tighter bound for $\V_N$ might be found using the aforementioned alternative controllability representation as per Assumption~\ref{asm:ADP-init-cost-bound}. 
\end{rem}

The following analysis merges the results of \cite{Koehler2021} and \cite{Heydari2018a} to ensure stability and deduce a performance mark on a level set of $\V_N$ in explicit dependence of the cost approximation error margin $c$ of $\eps_i$. 


\section{Stability and Performance Analysis}
\label{sec:analysis-main}

Let $\X_{\V}(\beta) \coloneqq \{x \in \X: \, \V_N(x) \leq \beta\}$ for $\beta>0$. 
The performance bound is given in the form
\begin{align} \label{eq:stability-performance}
	J(x(0),\kappa_{N}(\cdot)) \leq \frac{1}{\alpha_1(N,c)} \V_{N}(x(0)) \leq \frac{\alpha_2(N)}{\alpha_1(N,c)} V_{\infty}(x(0)),
\end{align}
for all $x(0) \in \X_{\V}(c)$, in which a $\underline{N} = \underline{N}(c) \in \N$ is sought that renders $\alpha_1(N,c) \in (0,1]$ for all $N \geq \underline{N}$ in dependence of the error bound $c>0$ of Assumption \ref{asm:ADP-error-bound} and which relates $\V_N$ to $V_{\infty}$ through $\alpha_2(N)$. 

In \eg \cite{Koehler2021} a similar description to \eqref{eq:stability-performance} has been pursued and the provided analysis can be transferred to the above setting under minor modifications. 
Using $\X_f$ of \eqref{eq:term-set}, the following results can be stated.

This description has been used in \cite{Koehler2021}, in which the analysis relates $\V_N$ to $l^\ast$ is various manners. 
Similarly, \cite{Heydari2018a} relates the cost $\hat{V}_i$ to $l^\ast$ (under the premise that $l^\ast(x) = l(x,0)$), which thus allows to merge the result to obtain the following. 

\begin{prp} \label{prp:alpha-1}
	Let Assumptions \ref{asm:stage-cost-lower-bound}--\ref{asm:ctrl-iter-input-constraint-satis} hold. 
	For any $\beta>0$ there exists $\underline{N}_1 = \underline{N}_1(c) \in \N$ \sut the first inequality of \eqref{eq:stability-performance} holds for all $x \in \X_{\V}(\beta)$ with $\alpha_1(N,c) >0$ and the closed-loop is asymptotically stable on $\X_{\V}(\beta)$ for all $N > \underline{N}_1$. 
	Specifically,
	\begin{align}
		\begin{split}
			&\underline{N}_1(c) \coloneqq \\
			&N' +  \frac{\max \{ 0, \ln(\underline{\gamma}_c) - \ln(1- c) ,\ln\left(  \frac{c(1- c) + 4 c \gamma_{0}}{(1-c)^2}\gamma_{\V} \right) \},}{ \ln(\gamma_{\V}) - \ln(\gamma_{\V} - 1)} 
		\end{split}
	\end{align}
where $\{N',\underline{\gamma}_c\}$ can be found in the proof in the appendix. 
\end{prp} 
Note that $\beta>0$ can be such that $\X_{\V}(\beta) \supset \X_f$.
Furthermore, the following holds. 

\begin{prp} \label{prp:alpha-2}
	Let Assumptions \ref{asm:stage-cost-lower-bound}--\ref{asm:ctrl-iter-input-constraint-satis} hold.
	For any $\beta>0$ there exists $\underline{N}_2 \in \N$ \sut the second inequality of \eqref{eq:stability-performance} holds for all $x \in \X_{\V}(\beta)$ with $\alpha_2(N) >0$ for all $N \geq \underline{N}_2$ and any $\alpha_1 >0$. 
\end{prp}

An explicit bound $\underline{N}_2$ can be found in the proof of Proposition~\ref{prp:alpha-2} in the appendix. 

Summarizing, it can be observed that for any $\beta>0$ there exists $\underline{N}(c) \in \N$ such that \eqref{eq:stability-performance} holds for all $x \in \X_{\V}(\beta)$ and all $N \geq \underline{N}(c)$. 
This horizon is given by $\underline{N}(c) \coloneqq \max \{\underline{N}_1(c),\underline{N}_2\}$. 
It may be noteworthy that $\lim_{N \ra \infty} J_{\infty}(x(0),\kappa_{N}(\cdot)) = V_{\infty}(x(0))$ from the proofs. 

The analysis concludes that there is a lower bound on the prediction horizon $N$ beyond which the predictive controller is asymptotically stabilizing and its performance \wrt to the optimal infinite-horizon cost can be related to the horizon length. 
The terminal cost, representing the optimal infinite-tail, can be trained offline on the nonlinear system and associated approximation errors in the sense of \eqref{eq:ADP-AVI-errors} can be accounted for directly via the prediction horizon to maintain stability and satisfy constraints. 
A controller tuning procedure is depicted by Algorithm \ref{alg:ctrl-tuning}.

\SetKwComment{Comment}{/* }{ */}
\begin{algorithm}
	\caption{Optimal cost approximation and predictive controller tuning.}	
	\label{alg:ctrl-tuning}
	\algrule
	\textbf{Preliminary offline computations:} \\
	\KwIn{dynamics $f(x,u)$; state cost $l(x,u)$; basis function vector $\Phi$; training set $\Omega$ represented by state samples $\hat{\Omega}$; test state samples $\hat{\Omega}_{\text{test}}$; local controller $\mu_{-1}$ on $\Omega$}
	\textbf{Controllability related calculations:} \\
	$\bullet$ get $(C,\sigma)$ of Asm. \ref{asm:loc-exp-controllability-iter-ctrl} on $\Omega$\;
	$\bullet$ find $d$ in $\X_f$ \sut condition below \eqref{eq:term-set} holds\;
	\textbf{Cost approximation related calculations:} \\
	$\bullet$ compute $w(0)$ of $\hat{V}_0$ in \eqref{eq:ADP-AVI-errors-1} using $\mu_{-1}$ (or set $\hat{V}_0 = V_0$ a Lyapunov function for $\mu_{-1}$, if known)\;
	$\bullet$ obtain $\gamma_{0}$ of Asm. \ref{asm:ADP-init-cost-bound}\;
	$\bullet$ get $c=c_{-1}  \geq 0$ of $|\eps_{-1}|$ in Asm. \ref{asm:ADP-error-bound}\;
set $i=0$\;
	\While{$i \leq I$ (or other abort criterion)}{
	$\bullet$ obtain $\mu_i$ in \eqref{eq:ADP-iter-ctrl}\;
	$\bullet$ compute $w(i+1)$ of $\hat{V}_{i+1}$ in \eqref{eq:ADP-AVI-errors-2}\; 
	$\bullet$ get $c =c_i \geq 0$ of $|\eps_{i}|$ in Asm. \ref{asm:ADP-error-bound}\; 
	}
	\eIf{$c_i \in [0,1)$ for all $i \in\{-1,\dotsc,\mathcal{I}\}$ and $\hat{\mu}(x) \in \U$ for all $x \in \Omega$}{take $c = \max_{i \in\{-1,\dotsc,\mathcal{I}\}} c_i$}{adjust $\Omega$ or/and $\Phi$, repeat}
	\KwOut{approximate optimal cost $\hat{V}$}
	\algrule
	\textbf{Predictive controller optimization settings:} \\
	\KwIn{cost $\hat{V}$; $(C,\sigma)$; $d$ of $\X_f$; error margin $c \in [0,1)$ from offline computations}
	$\bullet$ select $\beta>0$\;
	$\bullet$ compute $\underline{N}$ using $\{C,\sigma,d,\gamma_{0},c,\beta\}$ by Prop. \ref{prp:alpha-1} \& \ref{prp:alpha-2}\;
	$\bullet$ set $N \geq \underline{N}$ and $V_f = \hat{V}$\;
	$\bullet$ consider $x(0)$ such that $\V_N(x(0)) \leq \beta$. 
	\algrule
\end{algorithm}

The values $(C,\sigma)$ of the controllability assumption may be found using the training or test samples $\hat{\Omega}$ and simulating the closed-loop trajectory under $\mu_{-1}(\cdot)$. 
Computation of the value $d$ in $\X_f$ can be regarded as a maximization problem subject to set membership of $\X_f$ to $\Omega$. 
The value $\gamma_0$ can be obtained via $\hat{V}$ using the fact that $V_0(x) \leq 1/(1-c) \hat{V}_0(x)$ which has been pointed out in \cite[Lemma 1]{Heydari2018a}. 
Error margins for $\eps_i$, of which the maximum $c$ over all iterations is sought, are obtained by evaluating \eqref{eq:ADP-AVI-errors} at samples $x$ of $\hat{\Omega}_{\text{test}}$. 
Bound satisfaction for $\V_N(x(0))$ may be verified without optimization by using an arbitrary feasible $u(\cdot) \in \U^N$ in the cost of \eqref{eq:MPC-OCP}. 
If required, adjust $\beta$ and re-calibrate $N$ to catch the desired $x(0)$. 


\section{Case Study}
\label{sec:case-study}

This section demonstrates controller tuning along Algorithm \ref{alg:ctrl-tuning} and application of the resulting predictive controller on an orbital rendezvous problem that has been previously addressed in the context of approximate dynamic programming \cite{Heydari2014b}. 
The system's state given by $x(k) = [X(k),\, Y(k), \,X_t(k), \,Y_t(k)]^\top$ describes a spacecraft's position $[X,Y]^\top$ in two dimensions and their respective derivative $[X_t,Y_t]^\top$ \wrt time and follows the discretized dynamics
\begin{align*}
	\begin{split}
	&x(k+1) = \\
	&x(k) + \Delta t \left( \begin{bmatrix}
		X_t \\ Y_t \\ 2 Y_t - (1 + X) \left( \frac{1}{r^3} - 1\right) \\ -2 X_t - Y\left( \frac{1}{r^3} - 1 \right)
	\end{bmatrix} + \begin{bmatrix}
	 0 & 0 \\ 0 & 0 \\ 1 & 0 \\ 0 & 1
\end{bmatrix} u(k) \right)
\end{split}
\end{align*}
with $r = \sqrt{(1 + X)^2 + Y^2}$ and $\Delta t = 0.05$. 
In this study, let the stage cost be given by $l(x,u) = x^\top Q x + u^\top R u$, with $Q = \text{diag}(5,5,5,5)$ and $R = \text{diag}(1,1)$; a random choice. 
Consider $\X = [-0.5,0.5]^4$ and $\U = [-2 , 2]^2$; again, without specific consideration. 
The linearization of the system around the origin is locally controllable due to which a linear quadratic regulator can be constructed with linear control law $u_{\text{LQR}}(x) \coloneqq - K_{\text{LQR}} x$ and associated positive-definite cost matrix $P_{\text{LQR}}\succ 0$. 
Selecting $\mu_{-1}(\cdot) = u_{\text{LQR}}(\cdot)$, $(C,\sigma)$ of Assumption \ref{asm:loc-exp-controllability-iter-ctrl} can be computed via samples on $\Omega$-- take for instance $\Omega = [-0.2,0.2]^4$. 
Training of the approximate cost is performed using an approximant 
\begin{align}
	\label{eq:Vhat_case_study}
	\hat{V}_i(x) = w^\top(i) \Phi(x), \, 
\end{align}
comprising all monomials in the state variables of second and third degree. 
One may circumvent the initial cost computation based on $\mu_{-1}(x) = u_{\text{LQR}}(x)$ by selecting weights $w(0)$ such that $\hat{V}_0(x) = x^\top P_{\text{LQR}} x$. 
It has been pointed out in \cite[Lemma 1]{Heydari2018a} that $(1- c) V_0(x) \leq \hat{V}_0(x)$ from which $\gamma_0$ of Assumption \ref{asm:ADP-init-cost-bound} can be obtained by upper bounding $\hat{V}_0(x)$. 
\begin{figure}[h]
	\centering 
	\includegraphics[width = 0.9\columnwidth]{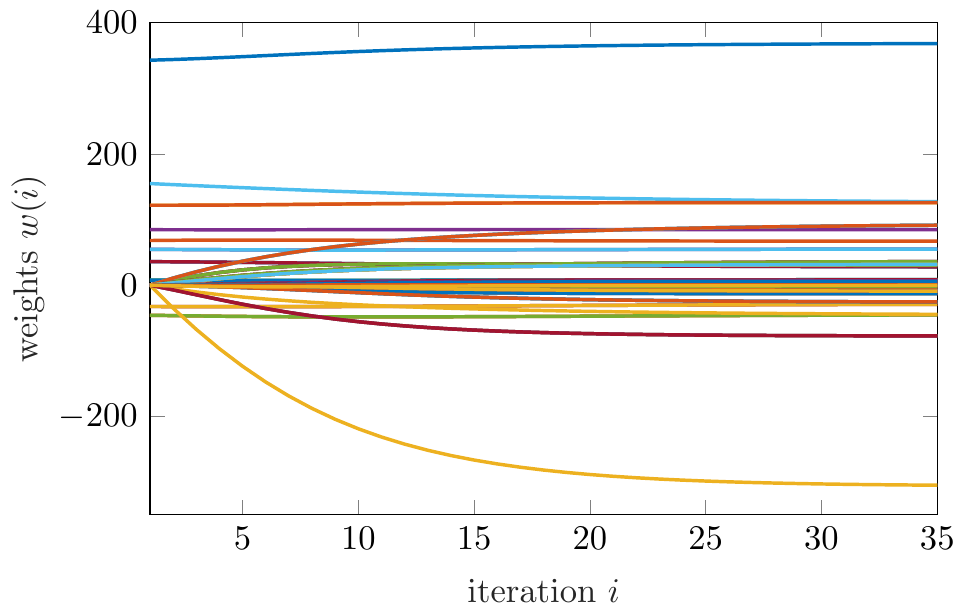}
	\caption{Converging weights of the optimal cost approximation.}
	\label{fig:AVI-offline}
\end{figure}
Fig.~\ref{fig:AVI-offline} shows the evolution of the weights $w(i)$ over $i$ under stabilizing value iteration on $p = 500$ samples $x^s \in \Omega $, $s = 1,\dotsc,p$. 
The weights converge after about $\mathcal{I} = 35$ iteration steps, leaving the approximant $\hat{V}(x) = w^\top(\mathcal{I})\Phi(x)$. 
Meanwhile, $\sup_{i \in \N_0,x^s \in \Omega} |\eps_{i}(x^s)| \leq 0.096$ with $c = 0.33$ -- for instance, the set $\Omega \in [-0.3,0.3]^4$ could not be covered by \eqref{eq:Vhat_case_study} with $c<1$. 
After convergence, an associated control law $\hat{\mu}(x)$ may be stated explicitly using \eg the approximation structure
\begin{align}
	\hat{\mu}(x) = w_a^\top \Phi_a(x), \; \; \Phi_a = [x^\top, (x \otimes x)^\top]^\top
\end{align}
to solve 
	$\hat{\mu}(x^s) = \mu_{\mathcal{I}}(x^s)$
for $w_a$ in a least-squares sense for samples $s = 1, \dotsc,p$. 
\begin{figure}[h]
	\centering
	\begin{subfigure}[t]{.49\columnwidth}
		\includegraphics[width=\columnwidth]{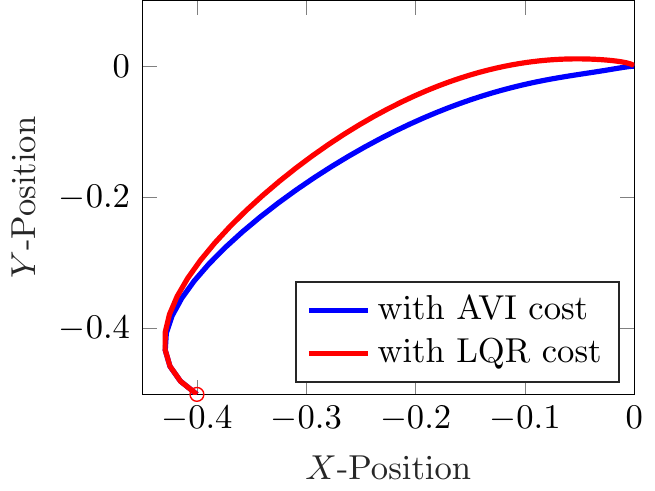}
	\end{subfigure} \hfill%
\begin{subfigure}[t]{0.49\columnwidth}
\includegraphics[width=\columnwidth]{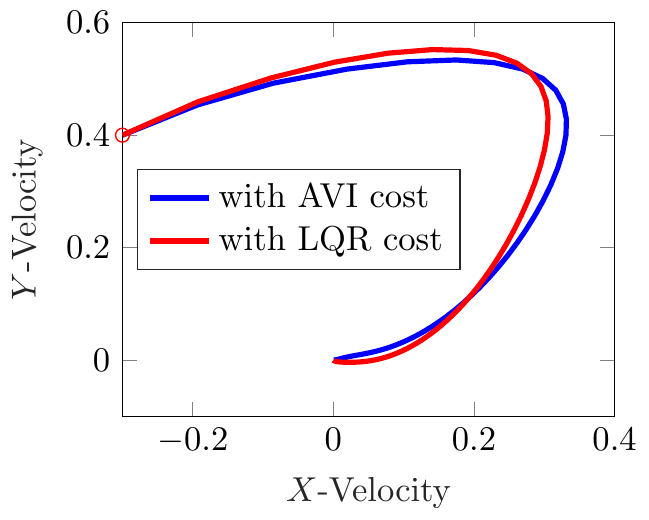}
\end{subfigure}
\caption{State trajectory under predictive control with terminal cost $V_f(x) = x^\top P_{\text{LQR}} x$ and $V_f(x) = \hat{V}(x)$.}
\label{fig:states}
\end{figure}
Fig.~\ref{fig:states} depicts the asymptotic behavior of the state towards the origin under different terminal costs $V_f$ configurations, using empirically $N = 10$. 
Due to the conservative bound of Proposition~\ref{prp:local-cost-bound} and values of $(C,\sigma) = (9.3204,0.9125)$, the actual required horizon $\underline{N}_1$ for decay is, unfortunately, considerably large -- more investigative work is required here.  


\section{Conclusion}

This work proposes an approximate infinite-horizon predictive controller tuning approach that constructs the terminal cost function of the controller optimization offline by means of approximate dynamic programming. 
The subsequent analysis relates the performance of predictive control to the reconstruction properties of the optimal cost function and provides a condition on the prediction horizon for asymptotic stability of the closed-loop. 
It is done by inspecting the iteration-step-wise errors in the approximation update, which is tractable by testing state samples (and continuity arguments). 
Depending on the prediction horizon, the region of attraction may be enlarged with regards to the approximation domain of interest and guarantees can be stated for the controller's behavior outside the training set of the approximants. 
Furthermore, different choices of the approximation architecture, which influence the approximation error, may lead to a favorable \ie shorter stabilizing prediction horizon. 
This work analyses one particular approximation scheme whereas other, including different error specifications, require individual investigation.


\appendix

\begin{proof}[of Prop. \ref{prp:alpha-1}]
	In its core, the proof resembles that of \cite[Thm.5, Part II]{Koehler2021}, however with $\eps = d/(2 \gamma_0 C)$ and $\gamma_{\V}$ substituted for $\gamma$. 
	Yet, dependence of the decay condition and required horizon length on the update error $c$ differ from the existing result and need certain adjustment. 
	The proof is thus briefly reviewed herein while differences are highlighted -- for more details, please refer to \cite{Koehler2021}. 
	
	First observe the lower boundedness as in $ \underline{\alpha}_l(\|x\|) \leq l^\ast(x) \leq \V_N(x)$ for all $x \in \X$. 
	By case distinction of whether $l^\ast(x) \leq \eps$, $\V_N(x) \leq \bar{\gamma}_c l^\ast(x) \leq \bar{\gamma}_c \overline{\alpha}_l(\|x\|)$ for all $x \in \X_{\V}(\beta)$, in which $\bar{\gamma}_c \coloneqq \max \{\gamma_{\V},\beta / \eps\}$. 
	Denote $u^\ast(\cdot,x)$ and $x_{u^\ast}(\cdot,x)$ the minimizing sequence to \eqref{eq:MPC-OCP} and the corresponding state along \eqref{eq:system}, respectively, for brevity. 
	Recall that $\kappa_{N}(x) = u^\ast(0,x)$. 
	From dynamic programming,
	\begin{align*} 
		\V_{N-i}(x_{u^\ast}(i,x)) = \V_N(x) - \sum_{k=0}^{i-1} l(x_{u^\ast}(k,x),u^\ast(k,x)),
	\end{align*}
	for any $i \in \{0,\dotsc,N\}$ and $x \in \X_{\V}(\beta)$. 
	Take $N'_{\text{real}} \coloneqq \max \{0 , (\beta - \gamma_{\V} \eps) / \eps\}$ and let $N' \in \N$ be \sut $\N' \geq N'_{\text{real}}$. 
	Following the steps of \cite{Koehler2021},  
it holds that
\begin{align}
	\label{eq:MPC-AVI-stability-proof-DP-bound-3}
	\begin{split}
		\V_{N-N}(x_{u^\ast}(N,x)) \leq  &\rho_{\gamma}^{N-N'}  \min\{ \gamma_{\V} l(x,\kappa_N(x)) , \underline{\gamma}_c \eps\},
	\end{split}
\end{align}
with $\underline{\gamma}_c \coloneqq \min \{\gamma_{\V},\beta/\eps\}$ and $\rho_{\gamma} = (\gamma_{\V} - 1) / \gamma_{\V}$.  

To make use of the local properties of $\hat{V}(x)$ on $\X_f$, it is required that $x_{u^\ast}(N,x) \in \X_f$ \ie
\begin{align*}
	l^\ast(x_{u^\ast}(N,x)) \leq &\frac{1}{1- c} \hat{V}(x_{u^\ast}(N,x)) \\
	= &\frac{1}{1- c} \V_{N-N}(x_{u^\ast}(N,x)) \leq \eps 
\end{align*}
in which the first inequality stems from i) of Thm. \ref{thm:ADP-AVI-Heydari}. 
Subsequently, 
\begin{align*}
	\frac{1}{1- c}\rho_{\gamma}^{N-N'}  \min\{ \gamma_{\V} l(x,\kappa_N(x)) , \underline{\gamma}_c \eps\} \leq \eps.
\end{align*}
is sought. 
Choosing
\begin{align}
	\label{eq:MPC-AVI-min-hoizon-term-state-set}
	N \geq N' + \frac{\max \{ 0, \ln(\underline{\gamma}_c) - \ln(1- c)  \}}{ \ln(\gamma_{\V}) - \ln(\gamma_{\V} - 1)} \eqqcolon \underline{N}'
\end{align}
gives the desired condition. 

Consider $\Delta \V_N(x) \coloneqq \V_N(f(x,\kappa_N(x))) - \V_N(x)$ for any $N \geq  \underline{N}'$, which, by optimality, gives 
\begin{align*}
	\Delta \V_N(x) \leq &-l(x,u^\ast(0,x)) + l(x_{u^\ast}(N,x),\tilde{u}) \\
	&+ V_f \left( f \left( x_{u^\ast}(N,x),\tilde{u} \right)  \right) - V_f \left( x_{u^\ast}(N,x) \right)
\end{align*}
for some feasible $\tilde{u} \in \U$, using $u^\ast(\cdot,x)$ as constraint admissible input at $f(x,\kappa_N(x))$.
Since $N \geq  \underline{N}'$, $x_{u^\ast}(N,x) \in \X_f$ and thus take $\tilde{u} = \hat{\mu}(x_{u^\ast}(N,x)) \in \U$. 
Then by \eqref{eq:Viter-difference-Heydari} for $V_f(x) = \hat{V}(x) = \hat{V}_{\mathcal{I}+1} (x)$, this is further bounded by 
\begin{align*}
	&\Delta \V_N(x) \\
	\leq &-l(x,\kappa_N(x))  - \eps_{I}(x_{u^\ast}(N,x)) + \frac{4 c}{1 - c} V_0(x_{u^\ast}(N,x)) \\
	\leq & -l(x,\kappa_N(x)) \\
	&+ c l^\ast(x_{u^\ast}(N,x)) + \frac{4 c}{1 - c} \gamma_{0} l^\ast(x_{u^\ast}(N,x)) \\
	\leq &-l(x,\kappa_N(x)) + \frac{ c(1- c)}{(1- c)^2} \hat{V}(x_{u^\ast}(N,x)) \\
	&+ \frac{4 c}{(1 - c)^2} \gamma_{0} \hat{V}(x_{u^\ast}(N,x))
\end{align*}
and by \eqref{eq:MPC-AVI-stability-proof-DP-bound-3}, 
\begin{align*}
	\Delta \V_N(x) \leq &-l(x,\kappa_N(x)) \\
	&+\rho_{\gamma}^{N-N'} \frac{c(1- c) + 4 c \gamma_{0}}{(1-c)^2} \gamma_{\V}l(x,\kappa_N(x)) \\
	= &- \underbrace{(1 - \rho_{\gamma}^{N-N'} \frac{c(1- c) + 4 c \gamma_{0}}{(1-c)^2}\gamma_{\V})}_{\coloneqq \alpha_1(N,c)} l(x,\kappa_N(x)). 
\end{align*}

A prediction horizon $N \in \N_0$ satisfying 
\begin{align*}
	N > N' + \frac{\ln\left(  \frac{c(1- c) + 4 c \gamma_{0} }{(1-c)^2}\gamma_{\V} \right)}{\ln(\gamma_{\V})  -\ln( \gamma_{\V} -1)}
\end{align*}
guarantees $\alpha_1(N,c) >0$ and hence negativity of $\Delta \V_N(x)$ for $x \in \X_V(\beta) \setminus \{0\}$. 
Subsequently, $\V_N(f(x,\kappa_N(x))) \leq  \V_N(x) \leq \beta$ for all $x \in \X_{\V}(\beta) $. 

Summarizing, the minimal horizon length is given by 
\begin{align}
	\begin{split}
	&\underline{N}_1(c) \coloneqq \\
	&N' +  \frac{\max \{ 0, \ln(\underline{\gamma}_c) - \ln(1- c) ,\ln\left(  \frac{c(1- c) + 4 c \gamma_{0}}{(1-c)^2}\gamma_{\V} \right) \},}{ \ln(\gamma_{\V}) - \ln(\gamma_{\V} - 1)} 
	\end{split}
\end{align}
and the closed-loop is asymptotically stable for all $N > \underline{N}_1(c)$.
	 
\end{proof}

\begin{proof}[of Prop. \ref{prp:alpha-2}]
	Denote $\eps \coloneqq d / 2 \gamma_{0}C$ and for brevity, $x^\ast(\cdot) $ the infinite-horizon optimal state trajectory starting at $x \in \X_{\V}(\beta)$ and $\kappa_{\infty}(\cdot)$ the associated optimal feedback to $V_{\infty}(x)$. 
	Recall that by Assumption~\ref{asm:loc-exp-controllability-iter-ctrl}, $V_{\infty}(x(0)) \leq \gamma l^\ast(x(0))$ for all $x(0) \in \X_f$ where $\gamma = C / (1- \sigma)$.
	
	Analogous to the proof of Prop. \ref{prp:alpha-1}, with $N''_{\text{real}} = \max\{0,\frac{c-\gamma \eps}{\eps}\}$ and $N'' \in \N_0$ \sut $ N''\geq N''_{\text{real}}$, 
%
%
one obtains
\begin{align*}
	V_{\infty}(x^\ast(N)) \leq & \left(\frac{\gamma - 1}{\gamma}\right)^{N-N''} V_{\infty}(x(0)).
\end{align*} 
as well as
	\begin{align*}
		V_{\infty}(x^\ast(N)) \leq & \left(\frac{\gamma - 1}{\gamma}\right)^{N-N''} \min \{ \gamma l (x, \kappa_{\infty}(x)), \underline{\gamma} \eps\},
	\end{align*}
	from which $l^\ast(x^\ast(N)) \leq V_{\infty}(x^\ast(N)) \leq \eps$ can be guaranteed for 
	\begin{align*}
		N \geq N'' + \frac{\max\{\ln(\underline{\gamma}),0\}}{\ln(\gamma) - \ln(\gamma-1)}  =: \underline{N}''.
	\end{align*}
	with $\underline{\gamma} \coloneqq \min \{\gamma,\beta/\eps\}$. 
	This is equivalent to \cite{Koehler2021}. 
	
	For any horizon $N \geq \underline{N}_2  \coloneqq \max\{\underline{N}',\underline{N}''\}$, with $\underline{N}'$ defined in \eqref{eq:MPC-AVI-min-hoizon-term-state-set}, it additionally holds that $\frac{1}{1-c} \V_{N-N}(x_{u^\ast}(N,x)) \leq \eps$ and thus $x_{u^\ast}(N,x) \in \X_f$ as well as $x^\ast(N) \in \X_f$. 
	Since therefore $\hat{V}(x^\ast(N)) \leq 2 \gamma_{0} l^\ast(x^\ast(N))$ by iii) of Thm. \ref{thm:ADP-AVI-Heydari}, it follows that
	\begin{align*}
		\hat{V}(x^\ast(N)) \leq &2 \gamma_{0} l^\ast(x^\ast(N)) \leq 2 \gamma_{0} V_{\infty}(x^\ast(N)) \\
		\leq &  2 \gamma_{0}  \left(\frac{\gamma- 1}{\gamma}\right)^{N-N''} V_{\infty}(x(0)).
	\end{align*}
	Because $x^\ast(k)$, $k \in \{0, \dotsc,N\}$, is a feasible candidate to \eqref{eq:MPC-OCP}, 
	\begin{align*}
		\V_N(x(0)) \leq &\sum_{k=0}^{N-1} l(x^\ast(k),\kappa_{\infty}(x^\ast(k))) + \hat{V}(x^\ast(N)) \\
		\leq &V(x)  + \hat{V}(x^\ast(N)) \\
		\leq &\underbrace{\left(1 + 2 \gamma_{0}  \left(\frac{\gamma - 1}{\gamma}\right)^{N-N''} \right)}_{ =: \alpha_2(N)} V_{\infty}(x(0)).
	\end{align*}
	for all $N \geq \underline{N}_2 $.
\end{proof}

\bibliographystyle{plain}
\bibliography{bib/MPC,bib/ClassicOptControl,bib/ADP_RL,bib/NonlinearControl}

\end{document}